\documentclass[a4paper,12pt]{article}
\usepackage{amsmath,amsfonts,amssymb,cite}
\usepackage{bm}

\begin{document}

\begin{center}
{\Large\bf Holographic like models as a five-dimensional rewriting
of large-$N_c$ QCD}
\end{center}

\begin{center}
{\large S. S. Afonin}
\end{center}

\begin{center}
{\it V. A. Fock Department of Theoretical Physics, Saint-Petersburg
State University, 1 ul. Ulyanovskaya, 198504, Russia}
\end{center}

\begin{abstract}
The AdS/QCD models are known to be tightly related with the QCD sum
rules in the large-$N_c$ (called also planar) limit. Rewriting the
theory of infinite tower of free stable mesons expected in the
large-$N_c$ QCD as a five-dimensional theory we scrutinize to what
extend the bottom-up holographic models may be viewed as an
alternative language expressing the phenomenology of planar QCD sum
rules. It is found that many features of AdS/QCD models can be
thereby obtained without invoking prescriptions from the original
AdS/CFT correspondence. Under some assumptions, all possibilities
leading to simple Regge trajectories are classified and it is argued
that the most phenomenologically consistent model is the one called
"soft wall model" in the holographic approach, with a preference to
the positive-sign dilaton background.
\end{abstract}

\newpage

\section{Motivation and formulation of problem}

In the last several years, numerous attempts to apply the ideas of
AdS/CFT correspondence~\cite{mald,witten} to QCD have formed a
rapidly developing line of researches in the modern field theory and
phenomenology. The main efforts are concentrated on description of
the hadron and glueball spectra, the chiral symmetry breaking, and
the related
phenomenology~\cite{pol,mateos,son0,sakai,kruc,son1,pom,br1,braga,par,katz,son2,andreev,eva,cs,gurs1,gurs2,hirn1,hirn2,kir,shock,casero,hong,kim,forkel,forkel2,krikun,fazio,ju,vega,pom2,gh,afonin,br2,br3,zuo,capp,bechi,sui,redi},
although the range of applications is now much broader. The
holographic methods are interesting as a language that allows to
discuss within a uniform framework various approaches to modeling
the interactions and spectral characteristics of light hadrons,
heavy-light systems, hadron form factors, QCD phase diagram, and
other phenomenological aspects which were previously the subjects of
investigations for different communities. This unifying property
brings up a natural question --- why the AdS/QCD models are rather
successful at reproducing the non-perturbative physics of strong
interactions and how the holographic language is interrelated with
other known phenomenological methods? At present there are plenty of
papers devoted to the applications of holographic ideas but there is
a lack of works which would try to answer this question\footnote{A
likely exception is the paper~\cite{redi} where it was demonstrated
numerically that the phenomenology of spin-1 mesons within the hard
wall holographic models depends slightly on the choice of metric,
the AdS metric, however, minimizes deviations from the experimental
data.}. The given research is intended to shed light on the problem
by observing that rewriting the large-$N_c$ QCD in a fixed channel
as a free five-dimensional theory, one is able to obtain models
which are very similar to the holographic models of the bottom-up
approach.

The AdS/QCD models are essentially based on the identification of 4D
Kaluza-Klein (KK) harmonics of a 5D field with an infinite tower of
colorless free meson states that is expected in the large-$N_c$
(called also planar) limit of QCD~\cite{hoof,wit}. Such an
identification can represent a convenient mathematical trick: Any
field $\varphi(\vec{x},z)$ living in a compact 5D space can be
decomposed in 4D harmonics,
$\varphi(\vec{x},z)=\sum_{n=0}^{\infty}\varphi_n(z)\psi_n(\vec{x})$,
which are 4D fields with masses determined by topology of the 5D
space. For this reason we may rewrite an infinite tower of 4D fields
$\psi_n(\vec{x})$ with equal quantum numbers and with a given
spectrum $m_n$ as a single 5D field $\varphi(\vec{x},z)$ propagating
in a suitable background. Decomposing this field in 4D harmonics
with normalizable coefficient functions $\varphi_n(z)$, the
dependence on the fifth coordinate $z$ can be integrated out and one
comes back to the original action for the tower of 4D fields
$\psi_n(\vec{x})$. The AdS/CFT conjecture, however, has much reacher
dynamical content than just KK-rewriting --- it asserts an exact
correspondence between the generating functional of correlators in
the 4D theory and the effective action of the 5D theory in which the
ultraviolet boundary values of the 5D fields are identified with the
sources in 4D theory~\cite{witten}. The question is whether we may
interpret the KK-rewriting in such a way that it mimicked the
correspondence above? The answer seems to be positive as is
demonstrated in Section~2. The idea is that after integration over
$z$ one obtains also boundary terms which may be identified with
sources coupled to the fields $\psi_n(\vec{x})$. This identification
yields the same results for the two-point correlators as within the
holographic models.

In Section~3, under some assumptions we analyze all possibilities
for the 5D backgrounds which lead to simple Regge like spectra and
then, in Section~4, we select the most reasonable 5D rewriting of
large-$N_c$ QCD basing on requirements to reproduce the analytic
structure  of two-point correlators and the expected
phenomenological form of Regge spectrum, this rewriting turns out to
be nothing but the so-called soft wall model~\cite{son2}.

The description of the chiral symmetry breaking is among the most
discussed issues in the holographic
approach~\cite{son1,pom,hirn1,hirn2,gh}. We propose in Section~5 a
modification of method of Refs.~\cite{son1,pom} that seems to allow
for embedding this phenomenon into our approach.

In Section~6, we consider interrelation between holographic
bottom-up approach and the phenomenology of QCD sum rules in the
large-$N_c$ limit.

In the concluding Section~7, we discuss briefly the relation of the
5D formalism to some traditional phenomenological approaches.

\section{Derivation of holographic like models via Kaluza-Klein reduction}

With respect to the number of colors $N_c$, the meson masses behave
as $m\sim N_c^0$ and their full decay widths do as $\Gamma\sim
N_c^{-1}$. Since the meson masses vary slightly with $N_c$, it is
thus often useful in the problems of finding the meson spectrum to
take the limit of large $N_c$. In the extreme case
$N_c\rightarrow\infty$, the mesons are infinitely narrow and
non-interacting, in addition, their appear in infinite
tower~\cite{hoof,wit} at fixed quantum numbers. Such an infinite
tower of resonance poles saturates completely the two-point
correlation function of quark currents with quantum numbers
corresponding to the given tower (see, e.g., the
representation~\eqref{68} in the vector channel). With the help of
functional integral formalism, this situation can be formally
described by the following action,
\begin{equation}
\label{1}
I_{[\mathcal{O}_J]}=(-1)^{J}\int d^4x\sum_{n=0}^{\infty}\left(\partial_{\mu}\phi_J^{(n)}\partial^{\mu}\phi^J_{(n)}-
m_{n,J}^2\phi_J^{(n)}\phi^J_{(n)}+\phi^J_{(n)}\mathcal{O}_J^{(n)}\right),
\end{equation}
where we have included the couplings to external sources which
appear in the corresponding generating functional of the connected
correlators. Here $\phi_J\doteq\phi_{\mu_1\mu_2\dots\mu_J}$,
$\mu_i=0,1,2,3$ with the signature $(+---)$, corresponds to a meson
field of spin $J$ whose quantum numbers $I,G,P,C$ are not specified,
and $\mathcal{O}_J^{(n)}$ is a source which can be represented as
\begin{equation}\label{1b}
\mathcal{O}_J^{(n)}=F_n^{(J)}\mathcal{O}_J,
\end{equation}
where $F_n^{(J)}$ are the decay constants defined by
\begin{equation}\label{1c}
\langle 0|\mathcal{O}_J^{(n)}|\phi_J^{(n)} \rangle=F_n^{(J)}\varepsilon_J
\end{equation}
for a meson $\phi_J^{(n)}$ with "polarization" $\varepsilon_J$ and
$\mathcal{O}_J$ is a common source to which the states
$\phi^J_{(n)}$ are coupled with a "coupling" $F_n^{(J)}$. In case of
high spin fields, $J>1$, under $\partial_{\mu}$ we mean the (general coordinate)
covariant derivative $\nabla_{\mu}$. Another subtlety is that for $J>1$ mesons
many additional quadratic terms appear in
the action~\eqref{1}, however, the auxiliary conditions can be chosen
such that they do not contribute (to be discussed below).
The representation~\eqref{1b} follows from the
requirement to obtain finally the standard form for two-point
correlators, namely integrating formally over the field $\phi_J$ in
the generating functional
\begin{equation}\label{1d}
Z[\mathcal{O}_J]=\int D \phi_J e^{I_{[\mathcal{O}_J]}},
\end{equation}
and differentiating twice with respect to $\mathcal{O}_J$ at
$\mathcal{O}_J=0$ one arrives at the sum over meson poles,
\begin{equation}\label{1e}
\left\langle\mathcal{O}_J(q)\mathcal{O}_J(-q)\right\rangle\sim\sum_{n=0}^{\infty}\frac{(F_n^{(J)})^2}{q^2-m_{n,J}^2+i\varepsilon},
\end{equation}
for the two-point correlators in the momentum representation. As is
usual in description of higher spin massive bosons, the tensor
$\phi_J$ is symmetric, traceless, $\phi^{\mu}_{\mu\dots}=0$, to
provide the irreducible $(\frac12 J,\frac12 J)$ representation of
the homogeneous Lorentz group (or, equivalently, the positivity of
energy), and satisfy the auxiliary condition
$\partial^{\mu}\phi_{\mu\dots}=0$ to give the required $2J+1$
physical degrees of freedom. The tensor structure in the r.h.s. of
representation~\eqref{1e} depends on the structure of terms with
derivatives in the action~\eqref{1}, it will be not of interest for
us.

We will not consider the baryons because their masses behave as
$m\sim N_c$ and their are not narrow at large-$N_c$. The baryons
emerge likely as solitonic objects in the large-$N_c$
limit~\cite{wit}, our further discussions are definitely not
applicable to such a case.

Our task is to rewrite the action~\eqref{1} as an action of some
free 5D theory that is not necessarily covariant with respect to the
fifth coordinate $z$, i.e. in the form
\begin{equation}\label{2}
S_{\text{5D}}=(-1)^{J}\int d^4x\,dz f_1(z)\left(\partial_M\varphi_J\partial^M\varphi^J-
m_J^2f_2(z)\varphi_J\varphi^J\right),
\end{equation}
where $M=0,1,2,3,4$, $\varphi_J=\varphi_J(x,z)$, and $f_1(z)$,
$f_2(z)$ are yet unknown functions of the fifth (space-like)
coordinate $z$. We require that the action~\eqref{2} must respect
the invariance under the 5D general coordinate transformations at
vanishing $|z|$. This implies two principal consequences. First,
\begin{equation}\label{3}
f_1(z)=e^{\Phi(z)}\sqrt{|\det{G_{MN}}|}=e^{\Phi(z)}a^5(z),
\end{equation}
where $G_{MN}$ is the metric tensor defined from the 5D metric,
\begin{equation}\label{4}
ds^2=a^2(z)(dx_{\mu}dx^{\mu}-dz^2),
\end{equation}
and possible deviations from the 5D covariance at large enough $|z|$
are pa\-ra\-met\-ri\-zed by means of a general factor $e^{\Phi(z)}$,
such a parametrization is convenient to draw parallels with the soft
wall AdS/QCD models. Second, one has to respect the rule for
contraction of indices, for instance,
\begin{equation}\label{5}
\partial_M\varphi_J\partial^M\varphi^J=\partial_M\varphi_{M_1\dots M_J}\partial_{M'}\varphi_{M_1'\dots M_J'}
G^{MM'}G^{M_1M_1'}\dots G^{M_JM_J'},
\end{equation}
where
\begin{equation}\label{6}
G^{MN}=G^{-1}_{MN}=a^{-2}(z)\eta^{MN}.
\end{equation}
We remind the reader again that $\partial_M$ denotes the covariant
derivative $\nabla_M$ for high spin fields. Finally, the action is
(up to a general normalization factor which is set to 1)
\begin{equation}\label{7}
S_{\text{5D}}=(-1)^{J}\int d^4x\,dz e^{\Phi(z)} a^{-2J+3}(z) \left\{(\partial_{\mu}\varphi_J)^2-(\partial_z\varphi_J)^2-
m_J^2a^2(z)\varphi_J^2\right\}.
\end{equation}
We will regard the 5D fields with $J>0$ as the gauge ones and use
this gauge freedom to go to such a gauge that no additional
quadratic terms appear in description of free higher-spin mesons.
For instance, in the case of AdS space (or at least the AdS space in
the UV limit), this is the axial gauge~\cite{son2,mikh},
\begin{equation}\label{8}
\varphi_{z\dots}=0,
\end{equation}
which will be implied in what follows. Together with the traceless
condition this gauge fixes components of fields $\varphi_J$.

Consider now the following Sturm-Liouville (SL) problem
\begin{equation}\label{9}
-\partial_z[g_J(z)\partial_z\varphi_n^{(J)}(z)]+g_J(z)a^2(z)m_J^2\varphi_n^{(J)}(z)=m_{n,J}^2g_J(z)\varphi_n^{(J)}(z).
\end{equation}
with
\begin{equation}\label{10}
g_J(z)=e^{\Phi(z)} a^{-2J+3}(z).
\end{equation}
We remind the reader the general results of SL theory in Appendix~A.
The equation~\eqref{9} is the equation of motion for the
action~\eqref{7} in the gauge~\eqref{8} if we admit
\begin{equation}\label{11}
\varphi(x,z)=e^{iq_nx}\varphi_n(z),\qquad q_n^2=m_n^2.
\end{equation}
The relation~\eqref{11} is usually interpreted as the string
excitation corresponding to 4D particle with physical momentum
$q_n$. {\it We will not use such an interpretation} and instead of
this assumption consider a general mathematical problem: Given a
spectrum $m_{n,J}^2$ in the action~\eqref{1}, find the functions
$g_J(z)$ and $a(z)$ which being inserted in~\eqref{9} would give
$m_{n,J}^2$ as eigenvalues of some SL problem. This constitutes the
main intermediate step in rewriting the action~\eqref{1} as some
free 5D field theory.

Under the conditions formulated in Appendix~A, the SL problem~\eqref{9} has the solutions $\varphi_n^{(J)}(z)$ which are normalized as follows
\begin{equation}\label{12}
\int_{z_{\text{min}}}^{z_{\text{max}}}g_J(z)\varphi_m^{(J)}(z)\varphi_n^{(J)}(z)dz=\delta_{mn},
\end{equation}
and form a complete set of functions, hence, the function $\varphi_J(x,z)$ in the action~\eqref{7} can be expanded
in the 4D harmonics,
\begin{equation}\label{13}
\varphi_J(x,z)=\sum_{n=0}^{\infty}\phi_J^{(n)}(x)\varphi_n^{(J)}(z).
\end{equation}
Let us substitute the expansion~\eqref{13} in the action~\eqref{7},
\begin{multline}
S_{\text{5D}}=(-1)^{J}\int d^4x\,dz g_J(z) \sum_{m,n=0}^{\infty} \left\{\varphi_m^{(J)}\varphi_n^{(J)}
\partial_{\mu}\phi_J^{(m)}\partial_{\mu}\phi_J^{(n)}\right.\\
\left.-\phi_J^{(m)}\phi_J^{(n)}\partial_z\varphi_m^{(J)}\partial_z\varphi_n^{(J)}
-m_J^2a^2(z)\varphi_m^{(J)}\varphi_n^{(J)}\phi_J^{(m)}\phi_J^{(n)}\right\},
\label{14}
\end{multline}
where the notation~\eqref{10} has been used. Integrating by parts
and making use of Eq.~\eqref{9}, the second term in the
action~\eqref{14} can be rewritten as (omitting the general factor
$-\phi_J^{(m)}\phi_J^{(n)}$)
\begin{multline}
\int_{z_{\text{min}}}^{z_{\text{max}}} dz g_J(z)\partial_z\varphi_m^{(J)}\partial_z\varphi_n^{(J)}=\\
\varphi_m^{(J)}g_J(z)\partial_z\varphi_n^{(J)}\left\vert_{z_{\text{min}}}^{z_{\text{max}}}\right.+
\int_{z_{\text{min}}}^{z_{\text{max}}} dz \varphi_m^{(J)}\varphi_n^{(J)}g_J(z)\left(m_{n,J}^2-a^2(z)m_J^2\right).
\label{15}
\end{multline}
Now we can integrate over $z$ in the action~\eqref{14} with the help of~\eqref{12} and~\eqref{15}, the result is
\begin{multline}
S_{\text{5D}}=(-1)^{J}\int d^4x\sum_{n=0}^{\infty}\left\{\left(\partial^{\mu}\phi^J_{(n)}\right)^2-
m_{n,J}^2\left(\phi^J_{(n)}\right)^2\right.\\
\left.-\phi^J_{(n)}g_J(z)\partial_z\varphi_n^{(J)}\varphi_J(x,z)\left\vert_{z_{\text{min}}}^{z_{\text{max}}}\right.\right\},
\label{16}
\end{multline}
with $\varphi_J(x,z)$ given by~\eqref{13}. The action~\eqref{16} has the form of the action~\eqref{1} if we identify
\begin{equation}\label{17}
\mathcal{O}_J^{(n)}=-g_J(z)\partial_z\varphi_n^{(J)}\varphi_J(x,z)\left\vert_{z_{\text{min}}}^{z_{\text{max}}}\right..
\end{equation}
If $\mathcal{O}_J^{(n)}=0$ then we have just rewritten the 4D action
through the 5D one as in the Kaluza-Klein reduction. But we want to
have an exact correspondence between the 4D action~\eqref{1} and the
5D action~\eqref{7}, i.e. $\mathcal{O}_J^{(n)}\neq0$. By choosing
$\beta=0$ or $\beta=\pi/2$ in the boundary condition~(A.3) (see
Appendix~A) we can nullify the term at $z=z_{\text{max}}$ in the
Eq.~\eqref{17}. Note in passing that for the vector mesons the first
possibility is actually realized in the soft wall holographic
models~\cite{son2}, where $z_{\text{max}}\rightarrow\infty$, while
the second one --- in the hard wall models~\cite{son1,pom,br1},
where $z_{\text{max}}$ is the infrared cutoff. Thus,
\begin{equation}\label{18}
\mathcal{O}_J^{(n)}=\lim_{z\rightarrow z_{\text{min}}+0} g_J(z)\partial_z\varphi_n^{(J)}(z)\varphi_J(x,z).
\end{equation}

Now we should identify the general source $\mathcal{O}_J$ such that
the decay constants $F_n^{(J)}$ in the representation~\eqref{1b}
were non-zero finite numbers. In general, this identification
depends on the choice of metric (the function $a(z)$) and of
five-dimensional mass $m_J^2$. If we fix the AdS metric at
$z\rightarrow z_{\text{min}}$, $a(z)\sim 1/z$, $z\geq0$, and $m_J^2$
in the form~\eqref{61} then we should identify
\begin{equation}
\mathcal{O}_J=zg_J\varphi_J/c\sim z^{2(J-1)}\varphi_J/c,
\end{equation}
\begin{equation}\label{19}
F_n^{(J)}=\lim_{z\rightarrow z_{\text{min}}+0}
c\frac{\partial_z\varphi_n^{(J)}(z)}{z},
\end{equation}
since the function $\varphi_n^{(J)}(z)$ in the realistic models
(Eqs.~\eqref{37} and~\eqref{40}) behave as $z^2$ at small $z$. The
factor $c$ appears because there is always the freedom to redefine
$F_n\rightarrow cF_n$, $\mathcal{O}\rightarrow \mathcal{O}/c$ where
the parameter $c$ can be fixed by matching to the high energy
asymptotics of the corresponding QCD correlators~\cite{co2}. The
expression~\eqref{19} reproduces the result of Ref.~\cite{son1} for
the vector mesons (up to a general factor depending on normalization
of fields).

Let us now compare the Kaluza-Klein like calculations above with the
technique used in the standard holographic approach. Within the
latter, one evaluates the action~\eqref{7} on the solution of
Eq.~\eqref{9} that would give in our notations
\begin{equation}\label{22b}
S_{\text{5D}}=(-1)^{J}\int d^4x\,g_J(z)\partial_z\varphi^J\varphi_J\left\vert_{z=z_{\text{min}}}\right..
\end{equation}
Supposing that $\varphi_0^J$ is the Fourier transform of the source
of QCD operator $\mathcal{O}^J$ at the UV boundary and accepting
$\varphi^J(q,z)=\varphi(q,z)\varphi_0^J(q)$ with the boundary
condition $z^{J-1}\varphi(q,z)\vert_{z\rightarrow +0}=1$ (we
generalize the matter to the case of arbitrary spin $J$) one
differentiates the Eq.~\eqref{22b} twice with respect to the source
and arrives at the same sum over meson poles as in the
expression~\eqref{1e} using the spectral expansion of the Green's
function for the Eq.~\eqref{9},
\begin{equation}\label{22c}
G^{(J)}(z,z';q)=\sum_{n=0}^{\infty}\frac{\varphi_n^{(J)}(z')\varphi_n^{(J)}(z)}{q^2-m_{n,J}^2+i\varepsilon}.
\end{equation}
This very coincidence enables the assumed holographic duality~\cite{witten}
\begin{equation}\label{22d}
Z_{\text{4D}}[\mathcal{O}_J]=S_{\text{5D}}[\varphi^J(x,z_{\text{min}})]
\quad \text{at} \quad
z_{\text{min}}^{J-1}\varphi^J(x,z_{\text{min}})=\mathcal{O}_J
\end{equation}
at the level of two-point correlation functions in the approximation of classical field theory.

It should be noted finally that we did not need the interpretation
of the fifth coordinate $z$ as inverse energy scale that is used in
the standard holographic approach.

\section{Regge spectrum}

On the theoretical and phenomenological~\cite{phen} grounds, the spectrum of light mesons is expected to be of the Regge form,
\begin{equation}m_{n,J}^2\simeq An+BJ+C,
\end{equation}
where $A$, $B$, $C$ are parameters, $J$ is the spin, and $n$ is the
"radial" quantum number that enumerates the states with identical
quantum numbers lying on the higher daughter trajectories. In
addition, one expects $A\approx B$, i.e. the behavior (see,
e.g.,~\cite{prc,sv} for the recent discussions)
\begin{equation}\label{23}
m_{n,J}^2\sim n+J.
\end{equation}
In the given Section, we examine systematically at which conditions the Regge spectrum can be obtained.

For further analysis it is convenient to cast the Eq.~\eqref{9} into
the form of the Schr\"{o}dinger equation via the substitution (we
omit below the index $(J)$ and the notation of dependence on $z$)
\begin{equation}\label{24}
\varphi_n=e^{-\Phi/2}a^{J-3/2}\psi_n.
\end{equation}
The result is
\begin{equation}\label{25}
-\psi_n''+U\psi_n=m_n^2\psi_n,
\end{equation}
\begin{equation}\label{25b}
U=\frac{\Phi''}{2}+\left(\frac{\Phi'}{2}\right)^2+
\left(\frac32-J\right)\dfrac{\Phi'a'+a''+\left(\frac12-J\right)\frac{(a')^2}{a}}{a}+a^2m_J^2,
\end{equation}
where the prime denotes the derivative with respect to $z$. As is
known from Quantum Mechanics, the linear dependence $m_n^2\sim n$
takes place if $U\sim z^2$ at least at large $z$, i.e. if one has
the oscillator type of potential. On the other hand, the Regge
behavior $m_{n,J}^2\sim J$ holds if there is a shift of energy
levels ($=$ masses square) $m_n^2\sim n+C(J)$ such that $C(J)\sim
J$. With the help of formulae in Appendix~B, the both conditions can
be achieved simultaneously at many choices of the functions
$\Phi(z)$ and $a(z)$ in the potential~\eqref{25b}. This freedom can
be restricted significantly if we make the following simplifying
assumptions:
\begin{description}
 \item {\bf(a)} The functions $\Phi$ and $a$, i.e. the shape of dilaton and the metric in the AdS/CFT language,
 do not depend on spin\footnote{Although there are speculations in the literature on that point~\cite{forkel}.} $J$;
 \item {\bf(b)} The functions $e^{\Phi}$ and $a$ are continuous and differentiable on the interval $(0,\infty)$ or $(-\infty,\infty)$
 (depending on a model);
 \item {\bf(c)} Subleading in $J$ and $n$ corrections to the Regge spectrum are absent\footnote{In reality, this seems to be not the case.
 The form of these corrections, however, is not known even approximately, it depends strongly on a model and is highly speculative in the phenomenology.
 We would not like to boil down into such speculations.}.
\end{description}

Under these assumptions we are able to classify all possible models yielding the Regge like spectrum.

The most important class of models that we will refer to as models~{\bf I} corresponds to the choice,

{\bf I}: $\Phi=\pm\lambda^2z^2$; $a=(R/z)^k$, where $0< z<\infty$.\\
The parameters $\lambda$ and $R$ have dimension of mass and inverse
mass correspondingly. Without loss of generality, we set $\lambda=1$
and $R=1$. The dependence on $\lambda$ and $R$ can be restored in
the final expressions by replacing $z\rightarrow \lambda z$,
$m_J\rightarrow m_JR^k$, $m_n\rightarrow m_n/\lambda$. The
equation~\eqref{25}-\eqref{25b} reads then as follows
\begin{multline}
-\psi_n''+\left\{z^2\pm\left[k\left(2J-3 \right)+1 \right] + \frac{\left[k\left(J-\frac32 \right) - \frac12 \right]^2- \frac14}{z^2} + \frac{m_J^2}{z^{2k}} \right\}\psi_n\\
=m_n^2\psi_n.
\label{25c}
\end{multline}
The normalizable solutions of Eq.~\eqref{25c} under different subcases are given below (see Appendix~B).

{\bf IA$^+$}: $\Phi=z^2$; $a=z^{-k}$, $k>1$; $m_J=0$.\\
The spectrum is
\begin{equation}\label{27}
m_n^2=4(n+1)+2(|\xi_{k,J}|+\xi_{k,J}),\qquad n=0,1,2,\dots,
\end{equation}
where
\begin{equation}\label{28}
\xi_{k,J}=k\left( J-\frac32 \right)-\frac12.
\end{equation}
The eigenfunctions (see the notation~\eqref{24})
\begin{equation}\label{29}
\varphi_n=\sqrt{\frac{2n!}{(|\xi_{k,J}|+n)!}}e^{-z^2}z^{|\xi_{k,J}|-\xi_{k,J}}L_n^{|\xi_{k,J}|}(z^2).
\end{equation}
The spectrum is absent for
\begin{equation}\label{30}
\frac32<J<\frac{1}{k}+\frac32.
\end{equation}
The restriction~\eqref{30} means that at $1<k<2$ there is no finite discrete spectrum for $J=2$ states. The case $k=2$ is special, the equation~\eqref{25c} is then
\begin{equation}\label{30b}
-\psi_n''+\left(z^2+3\right)\psi_n=m_n^2\psi_n,
\end{equation}
that yields formally the spectrum of masses and eigenfunctions given
by Eqs.~(B.4) and~(B.5). We must remember, however, that, first, due
to the pole in the metric the problem is defined in $0<z<\infty$
while the eigenfunctions~(B.5) are normalized in $-\infty<z<\infty$,
second, we should have $\varphi_n(0)=0$, i.e. only odd $n$ in~(B.5)
should be selected. Keepeng this in mind, we arrive at the spectrum
\begin{equation}\label{30c}
J,k=2:\qquad m_{2l+1}^2=4(l+3),\qquad l=0,1,2,\dots,
\end{equation}
and normalized eigenfunctions
\begin{equation}\label{30d}
J,k=2:\qquad\varphi_{2l+1}=\frac{\pi^{-1/4}}{2^{l+1}}\sqrt{\frac{1}{(2l+1)!}}e^{-z^2}z^{-1}H_{2l+1}(z).
\end{equation}

{\bf IA$^-$}: $\Phi=-z^2$; $a=z^{-k}$, $k>1$; $m_J=0$.\\
The equation~\eqref{25c} yields the spectrum
\begin{equation}\label{32}
m_n^2=4(n+1)+2(|\xi_{k,J}|-\xi_{k,J}),\qquad n=0,1,2,\dots,
\end{equation}
and the eigenfunctions
\begin{equation}\label{33}
\varphi_n=\sqrt{\frac{2n!}{(|\xi_{k,J}|+n)!}}z^{|\xi_{k,J}|-\xi_{k,J}}L_n^{|\xi_{k,J}|}(z^2).
\end{equation}
As before, the spectrum of $J=2$ mesons is absent for $1<k<2$ and for $k=2$ is given by
\begin{equation}\label{33b}
J,k=2:\qquad m_{2l+1}^2=4l,\qquad l=0,1,2,\dots,
\end{equation}
\begin{equation}\label{33c}
J,k=2:\qquad\varphi_{2l+1}=\frac{\pi^{-1/4}}{2^{l+1}}\sqrt{\frac{1}{(2l+1)!}}z^{-1}H_{2l+1}(z).
\end{equation}

{\bf IB$^+$}: $\Phi=z^2$; $a=1/z$.\\
The equation~\eqref{25c} is
\begin{equation}\label{34}
-\psi_n''+\left\{z^2 + 2(J-1) + \frac{(J-2)^2 + m_J^2 - 1/4}{z^2} \right\}\psi_n=m_n^2\psi_n.
\end{equation}
The spectrum
\begin{equation}\label{35}
m_n^2=4n+2(\xi_J+J),\qquad n=0,1,2,\dots,
\end{equation}
where
\begin{equation}\label{36}
\xi_J=\sqrt{(J-2)^2+m_J^2}.
\end{equation}
The eigenfunctions
\begin{equation}\label{37}
\varphi_n=\sqrt{\frac{2n!}{(\xi_J+n)!}}e^{-z^2}z^{\xi_J+2-J}L_n^{\xi_J}(z^2).
\end{equation}
The spectrum is absent for $0\leq\xi_J<1/2$ and represents a special
case for $\xi_J=1/2$ that can be analyzed as in the model~IA for any
concrete value of $m_J^2$.

{\bf IB$^-$}: $\Phi=-z^2$; $a=1/z$.\\
The equation~\eqref{25c} is
\begin{equation}\label{38}
-\psi_n''+\left\{z^2 - 2(J-1) + \frac{(J-2)^2 + m_J^2 - 1/4}{z^2} \right\}\psi_n=m_n^2\psi_n.
\end{equation}
The spectrum
\begin{equation}\label{39}
m_n^2=4(n+1)+2(\xi_J-J),\qquad n=0,1,2,\dots,
\end{equation}
\begin{equation}\label{40}
\varphi_n=\sqrt{\frac{2n!}{(\xi_J+n)!}}z^{\xi_J+2-J}L_n^{\xi_J}(z^2).
\end{equation}
As before, the spectrum is absent for $0\leq\xi_J<1/2$ and is special for $\xi_J=1/2$.

{\bf IC$^\pm$}: $\Phi=\pm z^2$; $a=z^{-k}$, $0<k<1$.\\
For $m_J=0$, this model is identical to the IA$^\pm$ one but with
one important distinction: The interval of forbidden
spins~\eqref{30} is now larger, in the limit $k\rightarrow+0$ only
the scalar and vector modes can be described. In the case
$m_J\neq0$, we are not aware of analytical solutions, presumably
they violate our assumption (c). It is clear, however, that at small
enough and large enough $z$ the solutions are approximately given by
the $m_J=0$ case.

{\bf ID$^\pm$}: $\Phi=\pm z^2$; $a=z^{-k}$, $k>1$; $m_J\neq0$.\\
The spectrum is given by
\begin{equation}\label{41}
m_n^2=4n\pm k(2J-3)+1+\dots,\qquad n=0,1,2,\dots,
\end{equation}
where the corrections presumably violate our assumption (c). The
eigenfunctions are not known, at large enough $z$ they are
approximately given by the model~IA$^\pm$.

The second class of models that we will denote the models~II corresponds formally to the limit $k=0$ in the models~I, namely

{\bf II}: $\Phi=\pm \lambda^2 z^2$; $a=\text{const}$, where $-\infty<z<\infty$.\\
Without loss of generality, we may set $\lambda=1$ and $a=1$. Note
that the absolute value of $m_J^2$ in the potential~\eqref{25b} is
not fixed because of freedom in choosing $a$. The model~II has two
different variants which are given below.

{\bf II$^+$}: $\Phi=z^2$.\\
The spectrum is
\begin{equation}\label{43}
m_n^2=2(n+1)+m_J^2, \qquad n=0,1,2,\dots,
\end{equation}
\begin{equation}\label{44}
\psi_n=\sqrt{\frac{1}{2^nn!}}\pi^{-1/4}e^{-z^2}H_n(z).
\end{equation}

{\bf II$^-$}: $\Phi=-z^2$.\\
The spectrum is
\begin{equation}\label{45}
m_n^2=2n+m_J^2, \qquad n=0,1,2,\dots,
\end{equation}
\begin{equation}\label{46}
\psi_n=\sqrt{\frac{1}{2^nn!}}\pi^{-1/4}H_n(z).
\end{equation}

There is the third, rather exotic class of models, it would
correspond to the choice $k=-1$ in the model~I but with a broader
possibility for the choice of function~$\Phi$,

{\bf III}: $a=z/R$.\\
The models~III can be divided into the following subcases.

{\bf IIIA$^{\pm}$}: $\Phi=\pm \lambda^2 z^2$; $\lambda^4+m_J^2/R^2>0$.\\
The equation~\eqref{25} is
\begin{equation}\label{48}
-\psi_n''+\left\{\left(\lambda^4+\frac{m_J^2}{R^2}\right)z^2 \pm 2\lambda^2(2-J) + \frac{(J-1)^2 - 1/4}{z^2} \right\}\psi_n=m_n^2\psi_n.
\end{equation}
The slope of Regge trajectories includes the 5D mass $m_J$, hence,
this quantity should not depend on $J$. Without loss of generality,
we may set $\lambda^4+m_J^2/R^2=1$ and $R=1$. The
equation~\eqref{48} does not describe the vector mesons, for
$J\neq1$ the spectrum is
\begin{equation}\label{49}
m_n^2=4n+2|J-1|+2\pm2\lambda^2(2-J) ,\qquad n=0,1,2,\dots,
\end{equation}
\begin{equation}\label{50}
\varphi_n=\sqrt{\frac{2n!}{(|J-1|+n)!}}e^{-\frac{1\pm\lambda^2}{2}z^2}z^{J-1+|J-1|}L_n^{|J-1|}(z^2).
\end{equation}
An  interesting particular case here is the choice $\lambda=0$ that
corresponds to the absence of dilaton background in the holographic
models, the slope of Regge trajectories is then determined by the
universal 5D mass square $m_J^2$.

{\bf IIIB}: $\Phi=b\log{(\lambda z)}$; $(m_J/R)^2>0$.\\
Choosing the units $\lambda=1$, the equation~\eqref{25} reads
\begin{equation}\label{51}
-\psi_n''+\left\{\left(\frac{m_J}{R}\right)^2z^2  + \frac{(J-1-b)^2 +b - 1/4}{z^2} \right\}\psi_n=m_n^2\psi_n.
\end{equation}
Here the slope is always determined by $m_J^2$. Setting for simplicity $(m_J/R)^2=1$, the spectrum is
\begin{equation}\label{52}
m_n^2=4n+2\xi_{J,b}+2 ,\qquad n=0,1,2,\dots,
\end{equation}
where
\begin{equation}\label{53}
\xi_{J,b}=\sqrt{(J-1-b)^2+b},
\end{equation}
and the eigenfunctions are given by
\begin{equation}\label{54}
\varphi_n=\sqrt{\frac{2n!}{(\xi_{J,b}+n)!}}e^{-z^2/2}z^{J-1+\xi_{J,b}-b/2}L_n^{\xi_{J,b}}(z^2).
\end{equation}
This model describes the mesons of all spins if
$b\geq(\sqrt{2}-1)/2$; at $b=(\sqrt{2}-1)/2$, the spectrum of $J=1$
mesons is given by Eqs.~(B.4) and~(B.5) with the same reservation as
for the model~IA.

\section{Choosing the most viable model}

After having classified all 5D models leading to simple Regge
trajectories we should impose some criteria which would select the
most viable model. Our first criterion is that a model should
describe the mesons of all spins in a uniform way, i.e. without
"pathologies" at certain values of $J$. On this ground, we reject
the models of type~IIIA and of type~IC.

The models in question can be matched to QCD by means of the
quark-hadron duality principle that is considered in Section~6. The
requirement of such a duality for the vector mesons constitutes our
second selection criterion. This  kind of matching for the present
models cannot be fulfilled for the $J>1$ mesons (at least we could
not find the relevant examples), this is not, however, a serious
drawback since the issue of quark-hadron duality for the $J\neq1$
mesons is questionable by itself\footnote{In the sense that the
partonic logarithm given by the free quark loop does not yield a
clear-cut main contribution to the two-point functions.}. Let us
show that the models of type~II do not pass this criterium.

Within the models under consideration, the condition of duality~\eqref{72} translates into
\begin{equation}\label{55}
\lim_{z\rightarrow0}\left[\partial_z\varphi_n(z) \right]^2\sim
m_n^2\partial_n m_n^2,\qquad n\gg1,
\end{equation}
where the l.h.s. is the residue in the case of type~II models and
$\varphi_n(z)\sim H_n(z)$ (see Eqs.~\eqref{44} and~\eqref{46}).
First of all, in order to have non-zero residues, we must define the
type~II models on the interval $0<z<\infty$, this changes slightly
the normalization of wave functions~\eqref{44} and~\eqref{46}. Due
to the identity
\begin{equation}\partial_z H_n(z)=2nH_{n-1}(z),
\end{equation}
only the states with uneven $n$ survive: $n=2l+1$, $l=0,1,2,\dots$. Making use of the property
\begin{equation}H_{2l}(0)=(-1)^l2^l(2l-1)!!,
\end{equation}
and the Regge form of the spectrum, $m_n^2\sim n$, we arrive at the expression
\begin{equation}\label{56}
\left[\partial_z \varphi_{2l+1}(0)\right]^2\sim m_{2l+1}^2h_l,\qquad h_l=\frac{\left[(2l-1)!!\right]^2}{(2l)!}.
\end{equation}
The quantity $h_l$ is a degreasing function of index $l$ while it
must be a constant (at least for large $l$) for the Regge like
spectrum. Thus, the condition of duality~\eqref{55} is not
fulfilled.

The condition~\eqref{55} is also very restrictive for the models of
type~I. First of all, it is easy to see that the residues are finite
and non-zero constants only if the 5D vector field is massless. In
this case (see Eq.~\eqref{28} and expressions for the relevant wave
functions)
\begin{multline}
\lim_{z\rightarrow0}g(z)\partial_z\varphi_n(z)\sim
\sqrt{\frac{n!}{(n+\frac{k+1}{2})!}}L_n^{\frac{k+1}{2}}(0)\sim
\sqrt{\frac{n!}{(n+\frac{k+1}{2})!}}\binom{n+\frac{k+1}{2}}{n}\\
\sim\sqrt{\frac{(n+\frac{k+1}{2})!}{n!}}.
\label{57}
\end{multline}
Substituting this expression in the duality condition~\eqref{55} we
conclude that the latter holds for $k=1$ only, i.e. for the AdS
background in the limit $z\rightarrow+0$. This situation corresponds
to the models of type~IB with $m_{J=1}^2=0$.

Our third selection criterion is the requirement to have in a
natural way the spectrum~\eqref{23}, i.e. the slopes of spin and
radial trajectories must coincide. We should emphasize a remarkable
fact that the third and the second criteria have a significant
overlap. For instance, the form of spectrum~\eqref{23} selects
immediately the models~IB among the models of type~I. Also it
excludes the model~IIIB. Thus, {\it the phenomenological
spectrum~\eqref{23} and the quark-hadron duality principle strongly
suggest that the AdS background in the limit $z\rightarrow+0$ is the
only possible 5D realization of large-$N_c$ QCD}.

We note further that the spectrum depends on spin $J$ not only by
construction (due to the covariant way for contraction of Lorentz
indices) but also through the dependence on $J$ of 5D mass $m_J$.
The latter dependence is {\it ad hoc} and requires invoking some
speculative assumptions. It would be natural therefore to have a
minimal impact of this dependence on the final results. In the
models of type~II, the spectrum depends on $J$ completely through
$m_J^2$, such a realization of Regge trajectories looks unfortunate
and for this reason it does not worth serious consideration.

The third criterion might be suggestive in the choice between the
models~IB$^+$ and~IB$^-$. In order to have a simple Regge-like
spectrum (our assumption (c))
\begin{equation}\label{58}
m_{n,J}^2=4(n+J+C),
\end{equation}
where $C$ is a constant and the value of the slope was determined by
the choice $\lambda=1$ in the dilaton background
$e^{\pm\lambda^2z^2}$, we must impose the following dependence of 5D
mass on spin $J$ (see Eqs.~\eqref{35},~\eqref{36}, and~\eqref{39})
\begin{align}
\label{59}
\text{IB}^+:&\qquad m_J^2=4J(C+1)+4(C^2-1),\\
\text{IB}^-:&\qquad m_J^2=8J^2+4J(3C-2)+4C(C-2).
\label{60}
\end{align}
The condition $m_{J=1}^2=0$ obtained above yields for the both cases
one acceptable solution $C=0$ that is, however, not valid for the
$J=0$ case in the model~IB$^-$. The five dimensional mass is then
\begin{align}
\label{61}
\text{IB}^+:&\qquad m_J^2=4(J-1),\\
\text{IB}^-:&\qquad m_J^2=8J(J-1),\qquad J>0.
\label{61b}
\end{align}
The both expression result in the spectrum
\begin{equation}\label{62}
m_{n,J}^2=4(n+J),\qquad J>0.
\end{equation}
The  spectrum~\eqref{62} turns out to be not valid for the scalars
of the model~IB$^+$ as well because the equation~\eqref{34} does not
have finite discrete spectrum in this case, i.e. when $m_{J=0}^2=-4$
according to Eq.~\eqref{61}. Thus the scalar mesons represent a
special case in the models~IB and should be considered separately.
It seems to be natural to supplement the relations~\eqref{61}
and~\eqref{61b} by the condition\footnote{Due to accidental
coincidence of this requirement with the Eq.~\eqref{61b} in the
$J=0$ case, the condition~\eqref{61c} may be imposed on the
model~IB$^+$ only.}.
\begin{equation}\label{61c}
m_{J=0}^2=0,
\end{equation}
The requirement~\eqref{61c} entails a change of
prescription~\eqref{19} for the residues, namely the latter must be
replaced for the scalar case by
\begin{equation}
F_n=\lim_{z\rightarrow z_{\text{min}}+0}
c\frac{\partial_z\varphi_n(z)}{z^3}.
\end{equation}
The condition~\eqref{61c} leads to the following spectrum for the
scalar mesons,
\begin{align}
\label{63}
\text{IB}^+:&\qquad m_{n,0}^2=4(n+1),\\
\text{IB}^-:&\qquad m_{n,0}^2=4(n+2),
\label{63b}
\end{align}
i.e. the scalars are degenerate with the vector mesons in the model~IB$^+$ and with the $J=2$ tensor mesons in the model~IB$^-$.

Consider now the impact of 5D mass $m_J$ on the final results of models~IB. For this purpose let us just set $m_J=0$, the spectra then are
\begin{align}
\label{63c1}
\text{IB}^+:&\qquad m_{n,J<2}^2=4(n+1),\\
\label{63c2}
            &\qquad m_{n,J\geq2}^2=4(n+J-1);\\
\label{63c3}
\text{IB}^-:&\qquad m_{n,J<2}^2=4(n+2-J),\\
\label{63c4}
            &\qquad m_{n,J\geq2}^2=4n.
\end{align}
We see that the impact of $m_J$ is rather slight in the model~IB$^+$
and is decisive in the model~IB$^-$. According to our third
criterion, the model~IB$^+$ looks therefore more attractive
phenomenologically than the model~IB$^-$. This conclusion, however,
may turn out to be misleading if we attempt to match the present
phenomenological approach to the theory of free higher spin fields
propagating in AdS$_5$ background where the mass coefficient behaves
as $J^2-J-4$ (see, e.g.,~\cite{mikh,buch} and references therein),
the related discussions are beyond the scope of the present paper.

Let us compare our results with the assumptions used in the
holographic models. According to the principle of AdS/CFT
correspondence, each operator $\mathcal{O}_J(x)$ in the 4D field
theory corresponds to a field $\varphi_J(x,z)$ in the 5D bulk theory
that has a mass $m_J$ determined via the relation~\cite{witten}
\begin{equation}\label{64}
m_J^2=(\Delta_J-J)(\Delta_J+J-4),
\end{equation}
where $\Delta_J$ is the canonical dimension of the operator
$\mathcal{O}_J(x)$. One can construct two types of minimal
interpolating operators which describe the mesons in QCD,
\begin{equation}\label{65}
\mathcal{O}_J^{t=2}=\bar{q}(\gamma_5)\gamma_{\{\mu_1}D_{\mu_2}\cdots D_{\mu_J\}}q,
\end{equation}
\begin{equation}\label{66}
\mathcal{O}_J^{t=3}=\bar{q}(\gamma_5)D_{\{\mu_1}D_{\mu_2}\cdots D_{\mu_J\}}q,
\end{equation}
where $t$ denotes the twist, $t=\Delta-J$. If we restrict ourselves
by the operators with minimal twist (a custom first approximation in
QCD) then $\Delta_J=J+2$, hence, according to Eq.~\eqref{64}, we
arrive exactly at the relation~\eqref{61}! The operators of the
minimal twist, however, cannot interpolate the scalar mesons, in
this case one uses the interpolator~\eqref{66} that predicts
$\Delta_0=3$, hence, $m_{J=0}^2=-3$. This does not contradict to our
analysis where the scalar case was also special. We would then
obtain for the model~IB$^+$
\begin{equation}\label{67}
m_{J=0}^2=-3:\qquad m_{n,0}^2=4(n+1/2).
\end{equation}
The phenomenology should tell us which spectrum of scalars is
preferable,~\eqref{63} or~\eqref{67}. If the spectrum~\eqref{63} is
definitely better phenomenologically then, within the holographic
approach, one should seemingly use the interpolator~\eqref{65} for
the scalars also, i.e. make contraction of indices and interpolate
the scalars by the twist-four operator
$\mathcal{O}_{J=0}^{t=4}=\bar{q}(\gamma_5)\gamma_{\mu}D_{\mu}q$,
thus $\Delta_0=4$ and $m_{J=0}^2=0$ according to Eq.~\eqref{64}. The
matter is however complicated by the fact that, firstly, this
interpolator is equivalent\footnote{Neglecting the axial anomaly in
the pseudoscalar isoscalar channel that is suppressed in the
large-$N_c$ limit.} to
$\mathcal{O}_{J=0}^{t=4}=m_q\bar{q}(\gamma_5)q$ due to the QCD
equations of motion and thus it can be neglected in the chiral
limit, $m_q=0$, secondly, one can construct a purely gluonic
twist-four interpolator $\mathcal{O}_{J=0}^{t=4}=\alpha_s
G_{\mu\nu}^2$ that makes difficult to distinguish between the
quark-antiquark states and glueballs in the scalar isoscalar channel
within the holographic approach.

The assumption of the AdS background in the limit $z\rightarrow0$
that is exploited in the AdS/QCD models matches perfectly our
results as was remarked above.

The spectrum~\eqref{62} was first obtained in the holographic model
of Ref.~\cite{son2}. We note that the sign of dilaton background was
opposite in that model (corresponding to the model~IB$^-$ in our
classification), this would lead to a different spectrum in our
approach which would coincide with the Eq.~\eqref{62} for the vector
mesons only (compare the Eqs.~\eqref{35} and~\eqref{39}). The reason
of this disagreement lies in a different way of introduction of
higher spin fields\footnote{We have introduced the higher spin
fields in a direct way as in paper~\cite{br2}. In the latter paper
(and analogous ones), however, the hadron spin is subsequently
devided into the orbital momentum of quark-antiquark pair and the
proper quark spin while we do not use such a model assumption.}
in~\cite{son2}. According to the arguments of recent
papers~\cite{br3,zuo} based on realization of confinement in the
string picture, the dilaton background should have positive sign in
the soft wall AdS/QCD models, i.e. the most consistent models are
those of type~IB$^+$ in our classification. We have arrived at the
same conclusion without using any string arguments.

\section{Chiral symmetry breaking}

The incorporation of the Chiral Symmetry Breaking (CSB) into 5D
approaches is often a questionable subject. Nevertheless we will try
to outline a possible description of the CSB phenomenon that is
compatible with our scheme of 5D holographic like rewriting.

First of all, since we have no quarks, no microscopic model of CSB
is possible, the best we may do is to describe the consequences of
CSB on the hadron level. The first such consequence is the
phenomenological fact that the masses of parity partners which
seemingly belong to the same chiral multiplet are quite different.
To reflect this effect one should introduce a mechanism for this
mass splitting. Within the 5D framework, the chiral symmetry does
not exist at all because there is no analogue for the matrix
$\gamma_5$ in five dimensions, the CSB can be therefore simulated
only indirectly by means of somewhat different description of states
with equal spin but opposite parity. We will restrict ourselves by
the vector and scalar sectors only.

The second consequence of CSB consists in the appearence of massless
(in the chiral limit) pseudoscalar meson due to the Goldstone
theorem. The both consequences should be described with the help of
a mechanism that explains, e.g., why the relation $m_{\pi}=0$ is
naturally related with $m_{a_1}^2\gtrsim 2m_{\rho}^2$ (instead of
$m_{a_1}=m_{\rho}$ as naively expected from the linear chiral
symmetry). The simplest such a mechanism that is realized in the
AdS/QCD models and that we will consider as well is borrowed from
the low-energy effective field theories: One introduces a scalar
field $X$ that acquires a non-zero vacuum expectation value (v.e.v.)
$X_0(z)$ and is coupled to the axial-vector field $A_{\mu}$ through
the covariant derivative~\cite{son1,pom},
\begin{equation}\label{82}
S_{\text{CSB}}=\int d^4 x\int dz \sqrt{|G_{MN}|}e^{\Phi(z)}\left(|D_M X|^2-m_X^2|X|^2\right),
\end{equation}
where
\begin{equation}\label{83}
D_M X=\partial_M X-ig_5A_MX.
\end{equation}
As usual, it is enough to restrict ourselves by the quadratic in
fields part in the action~\eqref{82} because the equations of motion
can provide a non-zero v.e.v. already in this case if the bulk space
is curved. This simplification turns out especially convenient for
us since it is much easier to integrate over $z$ in order to see the
equivalent 4D effective theory. The equation of motion determining
the v.e.v. $X_0$ is nothing but the SL equation~\eqref{9} for the
massless scalar particle. According to a recipe based on the AdS/CFT
correspondence~\cite{kleb}, it must behave at $z=0$ as
\begin{equation}\label{84}
X_0(z)\vert_{z\rightarrow0}=C_1z+C_2z^3,
\end{equation}
where $C_1$ is associated with the current quark mass, $C_1\sim
m_q$, and $C_2$ with the quark condensate,
$C_2\sim\langle\bar{q}q\rangle$. This interpretation implies a
somehow well established correspondence of the model to QCD. As we
do not have such a correspondence, it is more honest to say that the
incorporation of the 4D massless scalar particle can be related to
the spontaneous appearance of two order parameters with mass
dimension one and three and this property may be exploited to mimic
the CSB.

It is easy to establish a general self-consistency condition for
such a description of the CSB. Substituting $\varphi=X=z^h$ into the
Eq.~\eqref{9} with $a(z)=z^{-k}$ and zero r.h.s. ($k>0$, $h>0$) and
retaining the leading terms at $z\rightarrow0$, one has
\begin{equation}\label{85}
m_X^2z^{h-5k}=h(h-1-3k)z^{h-3k-2},
\end{equation}
that yields immediately $k=1$, i.e. {\it this design of CSB is
possible at $m_X^2\neq0$ only in the AdS background at
$z\rightarrow0$}. The second condition is $m_X^2=h(h-4)$ that both
for $h=1$ and for $h=3$ gives $m_X^2=-3$ in agreement with the
Eq.~\eqref{64} at $\Delta_0=3$. The case $m_X^2=0$ we will briefly
analyze later.

The fact that the pion does not belong to the corresponding linear
pseudoscalar trajectory and should be considered separately is in
agreement with the phenomenology. The solution $X_0(z)$, however,
turns out to be non-normalizable in the known bottom-up holographic
models. This is a serious problem for our approach because we cannot
integrate over $z$ and see the underlying 4D image of the
description. A way out may be the following. The equation of motion
for $X$ represents a second order linear differential equation that
has two independent solutions,
\begin{equation}\label{86}
X_0(z) = C_1X_1(z)+C_2X_2(z).
\end{equation}
The solution $X_1(z)$ spoils the normalizability at $z\rightarrow0$
while (in the soft wall models) $X_2(z)$ does at
$z\rightarrow\infty$. We do not see any reasons why $X_0(z)$ must
have everywhere a continuous derivative. Taking this into account,
we can construct the following normalizable solution,
\begin{equation}\label{87}
X_0(z) = C_2X_2(z)\vert_{z\leq z_0}+C_1X_1(z)\vert_{z>z_0}
\end{equation}
with the continuity condition
\begin{equation}\label{88}
C_1X_1(z_0)=C_2X_2(z_0).
\end{equation}
Another restriction on the input parameters comes from the normalization~\eqref{12},
\begin{equation}\label{88b}
C_2^2\int_{z_{\text{min}}}^{z_0}g_0(z)X_2^2(z)dz+C_1^2\int^{z_{\text{max}}}_{z_0}g_0(z)X_1^2(z)dz=1.
\end{equation}
The system of equations~\eqref{88},~\eqref{88b} allows to exclude two of three parameters $C_1$, $C_2$, $z_0$.

If $z_0$ is associated with the inverse energy scale then it is
natural to interpret $z_0^{-1}$ as the CSB scale,
$\Lambda_{\text{CSB}}\simeq4\pi f_{\pi}\approx1\div1.2$~GeV. The
physics of strong interactions is known to be substantially
different below $\Lambda_{\text{CSB}}$ and above
$\Lambda_{\text{CSB}}$, the effective emergence of this scale
represents in fact the third consequence of the CSB in QCD. This
consequence of the CSB seems to be not incorporated into the
existing AdS/QCD models.

Let us consider a concrete example --- the model IB$^+$ that has
been selected above as the most viable phenomenologically. The
corresponding solution for $X_0(z)$ is known~\cite{zuo},
\begin{equation}\label{89}
X_1(z)=ze^{-z^2}U(-1/2,0;z^2),
\end{equation}
\begin{equation}\label{90}
X_2(z)=z^3e^{-z^2}M(1/2,2;z^2),
\end{equation}
where $U$ and $M$ stay for the Kummer function $U$ and the Kummer function $M$ correspondingly. They have the asymptotics
\begin{equation}\label{91}
X_1(z)\xrightarrow[z\rightarrow0]{} z/\sqrt{\pi},\qquad X_1(z)\xrightarrow[z\rightarrow\infty]{} z^2e^{-z^2},
\end{equation}
\begin{equation}\label{92}
X_2(z)\xrightarrow[z\rightarrow0]{} z^3,\qquad X_2(z)\xrightarrow[z\rightarrow\infty]{} 1/\sqrt{\pi}.
\end{equation}
The continuity condition~\eqref{88} takes the form
\begin{equation}\label{93}
C_1 U(-1/2,0;z^2_0)=C_2 z_0^2 M(1/2,2;z^2_0),
\end{equation}
while the normalization~\eqref{88b} yields
\begin{equation}\label{94}
C_2^2\int_0^{z_0}z^3e^{-z^2}M^2(1/2,2;z^2)dz+C_1^2\int^{\infty}_{z_0}z^{-1}e^{-z^2}U^2(-1/2,0;z^2)dz=1.
\end{equation}

The solution of sum rules~\eqref{81} gives a remarkable relation for
the slope of meson trajectories, $a\simeq\Lambda_{\text{CSB}}^2$.
Identification of $z_0$ with $\Lambda_{\text{CSB}}^{-1}$ means then
in our units for $z$ that $z_0=1/2$. The system of
equations~\eqref{93},~\eqref{94} yields numerically
$C_1\approx\pm1.2$, $C_2\approx\pm3.5$.

The substitution of solution~\eqref{87} back into the
action~\eqref{82} gives the vacuum energy. As follows from the
asymptotics~\eqref{92}, the vacuum energy density
$\varepsilon_{\text{vac}}$ will be divergent logarithmically at
$z=0$,
\begin{equation}\label{94.b}
\varepsilon_{\text{vac}}=9C_2^2\log{z}\vert_{z\rightarrow0}+\text{const}.
\end{equation}
By subtracting the divergence in Eq.~\eqref{94.b}, we obtain the
renormalized vacuum energy density. In the case of numerical example
above it is
\begin{equation}\label{94.c}
\varepsilon_{\text{vac}}^{\text{(ren)}}\approx -29.3\vert_{z\leq0.5}+3.2\vert_{z>0.5}=-26.1.
\end{equation}
The negative value for $\varepsilon_{\text{vac}}^{\text{(ren)}}$ is in accord with expectations from the gluodynamics~\cite{migd}.

Comparing the prescription~\eqref{84} with the solution~\eqref{87}
and the asymptotics~\eqref{92}, we would say in AdS/QCD language
that the CSB is mimicked in the chiral limit (zero current quark
mass). This is self-consistent as long as we have assumed initially
the existence of exactly massless scalar particle.

Let us consider the case of vector mesons. First of all we note that
the trick above does not allow to introduce the massless vector
particles as bound states --- one can check that although
non-normalizable massless solutions exist for the spin-one mesons,
both of them are non-normalizable at $z\rightarrow0$. Thus, we are
save from the appearance of massless vectors\footnote{We would note
that there is a related logical uneasiness in some AdS/QCD models:
If one introduces a non-normalizable scalar field, why is this
forbidden in the vector channel?}.

The axial-vector sector is a traditional place for modelling the CSB
physics. We will not boil down to the model building of this kind
(the most popular way is the introduction of longitudinal component
for $A_M$ and of the pion field $\bm \pi$ through the
parametrization $X=X_0e^{i2{\bm \pi}}$) since in the chiral limit it
does not seam to bring new interesting results. An important
consequence of the term~\eqref{82} in the action is that the masses
of axial-vector mesons get shifted due to coupling to the v.e.v.
$X_0$ through the covariant derivative~\eqref{83}. Within the model
IB$^+$, the new mass spectrum can be in principle calculated via the
substitution $m_J^2\rightarrow m_J^2+g_5^2X_0^2(z)$ into the
Eq.~\eqref{34}. Although the exact analytic solution is not known,
it is clear that the masses will be enhanced because the new
contribution to the "potential" of Shr\"{o}dinger
equation~\eqref{34} is always positive. In addition, from the
solution~\eqref{87} and the asymptotics~\eqref{91},~\eqref{92}
follows that the holographic potential in the Eq.~\eqref{34} will be
(taking into account that $m_{J=1}^2=0$)
\begin{equation}\label{95}
U\vert_{z\gg z_0}=z^2\left(1+g_5^2C_1^2e^{-2z^2}\right)+\frac{3}{4z^2},
\end{equation}
\begin{equation}\label{96}
U\vert_{z\ll z_0}=z^2\left(1+g_5^2C_2^2z^2\right)+\frac{3}{4z^2}.
\end{equation}
This behavior demonstrates that the spectrum of axial-vector states
approaches rapidly to the spectrum of vector mesons and the
axial-vector residues are well-defined at least for high enough
excitations. The rate of such a "Chiral Symmetry Restoration" (CSR)
is exponential that is in agreement with the analysis~\cite{we,npb}
based on the QCD sum rules. This qualitative feature differs from
the previous results obtained within the soft wall holographic
models~\cite{gh,zuo} in which the shift between the $V$ and $A$
masses square tends to a constant and the axial-vector residues
cannot be determined from the expression~\eqref{19}. It must be
noted that the described behavior does not contradict to the CS
non-restoration scenario observed in the
phenomenology~\cite{prc,sv}. The latter states that the mesons on
the leading Regge trajectories do not have parity partners and the
spectral degeneracy has the form
\begin{equation}\label{97}
m_{n+1,J}^-\approx m_{n,J}^+\quad \text{($J$ uneven)};\qquad m_{n+1,J}^+\approx m_{n,J}^-\quad \text{($J$ even)},
\end{equation}
where $\pm$ refer to parity. The model above can be accommodated to such a scenario.

\section{QCD sum rules in the large-$N_c$ limit vs. AdS/QCD}

The momentum representation for the vector (V) and axial-vector (A)
two-point functions in the large-$N_c$ limit of QCD can be written
as (neglecting the nonpole terms)
\begin{equation}\label{68}
\left\langle
J_{\mu}^V(q)J_{\nu}^V(-q)\right\rangle=\Pi_{\mu\nu}^{\perp}(q)\sum_{n=0}^{\infty}\frac{F_{V,n}^2}{q^2-m_{V,n}^2+i\varepsilon},
\end{equation}
\begin{equation}\label{69}
\left\langle
J_{\mu}^A(q)J_{\nu}^A(-q)\right\rangle=q_{\mu}q_{\nu}\frac{f_{\pi}^2}{q^2}+
\Pi_{\mu\nu}^{\perp}(q)\sum_{n=0}^{\infty}\frac{F_{A,n}^2}{q^2-m_{A,n}^2+i\varepsilon},
\end{equation}
where $F_n$ are the corresponding electromagnetic meson decay constants, $f_{\pi}$ denotes the pion weak decay constant, and
\begin{equation}\label{70}
\Pi_{\mu\nu}^{\perp}(q)=-\eta_{\mu\nu}+\frac{q_{\mu}q_{\nu}}{q^2}
\end{equation}
is the transverse projector.

On the other hand, one can write the Operator Product Expansion (OPE) for those two-point functions at large Euclidean momentum, $Q^2=-q^2$,
\begin{equation}\label{71}
\left\langle J_{\mu}(q)J_{\nu}(-q)\right\rangle=\Pi_{\mu\nu}^{\perp}(q)Q^2
\left(C_0\ln\!\frac{\Lambda^2}{Q^2}+\sum_{i=1}^{\infty}\frac{C_i}{Q^{2i}}\right).
\end{equation}
Here $\Lambda$ is a renormalization scale and the coefficients $C_i$
depend weakly on $\Lambda$ and $Q^2$. It is easy to show that in
order to obtain the analytic behavior~\eqref{71} the residues in
Eqs.~\eqref{68} and~\eqref{69} must satisfy the following condition
(see, e.g.,~\cite{we} for a detailed discussion)
\begin{equation}\label{72}
F_n^2\sim m_n^2\partial_n m_n^2.
\end{equation}
Thus, given the relation~\eqref{72}, the infinite number of meson
poles is equivalent to the partonic logarithm in the Eq.~\eqref{71}
plus some corrections. This property we refer to as the quark-hadron
duality in the large-$N_c$ limit (brief reviews can be found
in~\cite{we,npb,beane}). Matching the Eq.~\eqref{71} to the
representations~\eqref{68} and~\eqref{69} in the Euclidean region,
one obtains the QCD sum rules which represent certain relations on
the parameters of the meson spectrum~\cite{we,npb,beane,sr,phil}.

Consider the simplest example --- the linear spectrum
\begin{equation}\label{73}
m_{V\!,A;n}^2=a(n+b_{V\!,A;n}),
\end{equation}
where the slope $a$ is assumed to be universal. According to the condition~\eqref{72}, the residues must behave as
(we neglect possible corrections to the relation~\eqref{72}~\cite{we})
\begin{equation}\label{73b}
F_{V\!,A;n}^2\sim m_{V\!,A;n}^2 F_{V\!,A}^2,\qquad F_{V\!,A}^2=\text{const}.
\end{equation}
Performing the summation in $n$ in the expression~\eqref{68}, one arrives at~\cite{npb,phil}
\begin{multline}
\label{74}
\left\langle J_{\mu}^V(q)J_{\nu}^V(-q)\right\rangle=\\
\Pi_{\mu\nu}^{\perp}(q)Q^2\frac{2F_V^2}{a}
\left[\ln\!\frac{a}{Q^2}+\sum_{i=1}^{\infty}\frac{(-1)^ia^iB_i(b_V)}{iQ^{2i}}+\text{const}\right],
\end{multline}
where $B_i(x)$ are the Bernoulli polynomials. The OPE for the two-point functions at $N_c\gg1$ and in the chiral limit reads as follows~\cite{svz}
\begin{multline}
\label{75}
\left\langle J_{\mu}^{V\!,A}(q)J_{\nu}^{V\!,A}(-q)\right\rangle=\\
\Pi_{\mu\nu}^{\perp}(q)Q^2\left[
\frac{N_c}{12\pi^2}\left(1+\frac{\alpha_s}{\pi}\right)\ln\!\frac{\Lambda^2}{Q^2}
+\frac{\alpha_s}{12\pi}\frac{\langle G^2\rangle}{Q^4}
+\frac{4\xi^{V\!,A}}{9}\pi\alpha_s\frac{\langle\bar{q}q\rangle^2}{Q^6}
+\mathcal{O}\left(\frac{1}{Q^8}\right)\right].
\end{multline}
Here $\langle G^2\rangle$ and $\langle\bar{q}q\rangle$ denote the
gluon and quark condensates and $\xi^V=-7$, $\xi^A=11$. Comparing
the Eq.~\eqref{75} with~\eqref{74} and with analogous expression for
the axial-vector two-point function, one obtains the following sum
rules (we consider the leading order in $\alpha_s$),
\begin{eqnarray}\label{76}
F_V^2=F_A^2\equiv F^2&=&\frac{N_ca}{24\pi^2},\\
\label{77}
0&=&b_V-\frac12,\\
\label{78}
\frac{f_{\pi}^2}{F^2}&=&b_A-\frac12,\\
\label{79}
\frac{\alpha_s\langle G^2\rangle}{12\pi F^2a}&=&b_{V\!,A}^2-b_{V\!,A}+\frac16,\\
\label{80}
-\frac{2\xi^{V\!,A}\pi\alpha_s\langle\bar{q}q\rangle^2}{3F^2a^2}&=&
b_{V\!,A}\left(b_{V\!,A}-1/2\right)\left(b_{V\!,A}-1\right).
\end{eqnarray}
The sum rules~\eqref{76}-\eqref{80} cannot be satisfied by the
simple linear spectrum~\eqref{73}, a natural solution of the problem
is to consider nonlinear corrections to the ansatz~\eqref{73} or/and
not place some states on the linear trajectory. We note, however,
that the sum
rules~\eqref{76}--\eqref{77},\eqref{78}--\eqref{79}--\eqref{80}--$\dots$
are progressively less and less reliable due to the asymptotic
nature of OPE and growing anomalous dimensions of condensate terms.
The simple linear spectrum is able to satisfy the most reliable
relations~\eqref{76}-\eqref{78}. Identifying $F\equiv F_{\rho}$ and
making use of the KSFR relation~\cite{ksfr} $F_{\rho}^2=2f_{\pi}^2$
as in the original Weinberg sum rules~\cite{wein}, we have
immediately the solution
\begin{equation}\label{81}
b_V=1/2,\qquad b_A=1,\qquad a=48\pi^2f_{\pi}^2/N_c,
\end{equation}
which is in a good agreement with the phenomenology (for instance,
the Weinberg relation $m_{a_1}^2=2m_{\rho}^2$ follows
automatically). This solution for the intercept $b_{V\!,A}$ yields
the spectrum of the generalized Lovelace-Shapiro (LS) dual
amplitude~\cite{avw}. The difference $b_A-b_V$ emerges from the CSB
dynamics that is directly seen in the Eq.~\eqref{78} (namely the
fact that $f_{\pi}\neq0$). Within the LS amplitude, the same
difference arises from taking into account the Adler
self-consistency condition~\cite{adler} which is a logical
consequence of the Goldstone theorem.

Turning back to the holographic approach, we note that the
spectrum~\eqref{62} that often appears in the soft-wall AdS/QCD
models, in fact, does not describe the $\rho$ and $\omega$ mesons as
opposed to the usual belief, it corresponds rather to the
axial-vector mesons. We are not aware of any natural way for
accommodation of the intercept $b_V=1/2$ within the holographic
approach and/or the technique presented in the previous sections,
unless some speculative extra assumptions are
involved~\cite{forkel,br3}. As follows from the discussions above,
the correct vector spectrum should naturally stem from a successful
implementation of the CSB physics in the holographic
models\footnote{The situation resembles the almost 40 years old
discussions about a still unresolved problem: In contrast to the
Veneziano dual amplitude that gave rise to the modern string theory,
the LS amplitude does not ensue from any string model; the
incorporation of the CSB effect into the string approach is a
formidable task up to now.}.

As seen from the expansion~\eqref{74}, the spectrum of masses (at
least in the vector channels) determines the gauge-invariant QCD
condensates; if there are several such condensates of equal
dimension it determines a certain linear combination of them. For
this reason, any attempts of inclusion of QCD condensates into the
soft wall holographic models face a serious problem of double
counting. This remark, however, is not relevant for the hard wall
models~\cite{son1,pom,br1}, where the masses are given by roots of
Bessel function, $m_n\sim n$, and for this reason it turns out that
the corrections to the partonic logarithm does not reproduce the
analytic structure of OPE.

Our previous discussions incline us to conclude that the bottom-up
holographic approach may represent just an alternative language for
expressing the phenomenology known from the QCD sum rules in the
large-$N_c$ limit with practically the same number of input
parameters and ensuing accuracy. This is directly seen from the
bottom-up derivation of AdS/QCD like models in Section~2 and {\it a
posteriori} from the type of questions which are addressed within
the holographic models. First of all, the problem of finding the
fittest spectrum in the sum rules is reformulated within the AdS/QCD
models as the problem of finding an appropriate 5D background and
boundary conditions on the 5D fields which give the required
spectrum as a result of solution of the corresponding SL problem
while in the sum rules the spectrum is looked for empirically
ignoring its model-dependent origin. The subsequent phenomenology,
however, is very similar in the both approaches. For instance, both
sum rules and most of holographic models incorporate the pion via
the Partial Conservation of Axial Current hypothesis. As is known
from the phenomenology of sum rules~\cite{we}, it is difficult to
accommodate the pion as the first state in the tower of pseudoscalar
resonances, one should rather consider it separately as a special
state, the same conclusion seems to follow from the AdS/QCD models.

\section{Concluding remarks}

The phenomenology expected in the large-$N_c$ limit of QCD can be
compactly rewritten in terms of some phenomenological 5D theory,
namely the infinite towers of narrow meson states coupled to
external sources can be formally represented as some 5D single
fields propagating in a suitable background. This allows to replace
sometimes lengthy manipulations with infinite series of meson poles
by compact operations with those 5D fields. Needless to say that
such an introduction of 5D formalism represents a certain
operational progress and one could even draw a parallel with the
introduction of continual integral into the field theory in the
sense that manipulations in the latter formalism are usually
understood as a compact formal way to write lengthy operations with
series of perturbation theory. The modern 5D model building,
however, involves various speculative assumptions borrowed from the
AdS/CFT correspondence which look often unfounded in their
extrapolation to real QCD. In the present work, we have explicitly
demonstrated that 5D models very similar to the so-called soft wall
holographic models can be derived within the 5D formalism imposing
some requirements of phenomenological consistency without use of
AdS/CFT dictionary. These models represent just another language for
expressing the Regge phenomenology within the framework of QCD sum
rules.

The 5D formalism proposes new ways for description of the chiral
symmetry breaking. The related attempts cannot be viewed as another
language for expressing the matching of low-energy effective field
theories for strong interactions to QCD sum rules (see,
e.g.,~\cite{effmod} among others), but are similar in spirit though.
The reason is that the effective models deal with the lowest mass
(the first radial excitation at best) mesons while the 5D models are
able to describe the impact of CSB on the whole tower of excited
states. Another reason consists in the fact that the existing 5D
descriptions of CSB are vague from the 4D point of view. We have
tried to advance in "visualizing" a 5D description of CSB in 4D
terms but could not provide a concrete example. The main problem to
be solved is the derivation of GOR relation for the pion mass within
the presented approach.

We did not discuss an explicit incorporation of quarks and QCD
running coupling since without knowledge of the confinement
mechanism, at least in the planar limit of QCD, any such attempt is
doomed to be rather speculative.

It must be emphasized that the considered 5D approach is completely
phenomenological and hardly can be directly related to any
fundamental extra-dimensional theory. First of all, the 5D
background of the soft wall model that we have also obtained cannot
be a solution of 5D Einstein equations as it breaks explicitly the
5D general covariance. A possible way out could consist in
constructing an alternative mechanism for mass generation. Such an
attempt for flat metric (an "effective" holographic model) was
undertaken in Ref.~\cite{egm1} (see also~\cite{egm2}) where the
Higgs mechanism was used. Second, we have ignored all long-standing
problems in description of higher spin fields. If we included a
gravitational dynamics that would yield particular metrics in
question we would have the problem of gravitational coupling for
higher spin fields --- such a theory does not exist. For this reason
we believe that within the present approach it is meaningless to
analyze the backreaction of metric due to condensation of scalar
field when modeling the CSB physics as well as to address other
questions of relation to consistent extra-dimensional theories until
some fundamental problems in field theory are solved. Further
discussions of relevant problems can be found in~\cite{egm3}.

\section*{Acknowledgments}

I am grateful for the warm hospitality by Prof. Maxim Polyakov
extended to me at the Department of Theoretical Physics of Ruhr
University Bochum and to all colleagues visiting or working at the
Department who inspired me to study the holographic approach to QCD
and whose critical attitude to this approach shaped partly the
content of the present research. The work is supported by the
Alexander von Humboldt Foundation and by RFBR, grant 09-02-00073-a.

\section*{Appendix A: Sturm-Liouville theorem}

In this Appendix, we briefly remind the reader the main results of the Sturm-Liouville (SL) theory. The SL equation is
\begin{equation*}
-\partial_z[p(z)\partial_z\varphi]+q(z)\varphi=\lambda\omega(z)\varphi.
\eqno{\text{(A.1)}}
\end{equation*}
Here the function $p(z)>0$ has a continuous derivative, the
functions $q(z)>0$ and $\omega(z)>0$ are continuous on the finite
closed interval $[z_{\text{min}},z_{\text{max}}]$. Loosely speaking,
the SL problem consists in finding the values of $\lambda$ for which
there exists a non-trivial solution of Eq.~(A.1) satisfying certain
boundary conditions. Under the assumptions that $p(z)^{-1}$, $q(z)$,
and $\omega(z)$ are real-valued integrable functions over the
interval $[z_{\text{min}},z_{\text{max}}]$, with the boundary
conditions of the form
\begin{equation*}
\varphi(z_{\text{min}})\cos\alpha-p(z_{\text{min}})\varphi'(z_{\text{min}})\sin\alpha=0,
\eqno{\text{(A.2)}}
\end{equation*}
\begin{equation*}
\varphi(z_{\text{max}})\cos\beta-p(z_{\text{max}})\varphi'(z_{\text{max}})\sin\beta=0,
\eqno{\text{(A.3)}}
\end{equation*}
where $\alpha,\beta\in[0,\pi)$ and prime denotes the derivative, the SL theorem states that
\begin{itemize}
\item There is infinite discrete set of real eigenvalues $\lambda_n$, $n=0,1,2,\dots$.
\item Up to a normalization constant, there is a unique eigenfunction $\varphi_n(z)$
corresponding to each eigenvalue $\lambda_n$ and this eigenfunction has exactly $n-1$ zeros in $[z_{\text{min}},z_{\text{max}}]$.
\item The normalized eigenfunctions form an orthonormal basis
\begin{equation*}
\int_{z_{\text{min}}}^{z_{\text{max}}}\varphi_m(z)\varphi_n(z)\omega(z)dz=\delta_{mn}.
\eqno{\text{(A.4)}}
\end{equation*}
\end{itemize}
Thus the solutions of the SL problem form a complete set of
functions in the interval $[z_{\text{min}},z_{\text{max}}]$, i.e.
they can be used for expansion of arbitrary functions in that
interval.

\section*{Appendix B: Schr\"{o}dinger equations appearing in the work}

The spectrum is typically determined by the equation
\begin{equation*}
-\psi_n''+\left[z^2+\frac{b^2-1/4}{z^2}+c\right]\psi_n=m_n^2\psi_n,
\eqno{\text{(B.1)}}
\end{equation*}
that is exactly solvable. The spectrum of normalizable modes is
\begin{equation*}
m_n^2=4n+2|b|+2+c, \qquad n=0,1,2,\dots,
\eqno{\text{(B.2)}}
\end{equation*}
the corresponding normalized eigenfunctions are
\begin{equation*}
\psi_n=\sqrt{\frac{2n!}{(|b|+n)!}}e^{-z^2/2}z^{|b|+1/2}L_n^{|b|}(z^2),
\eqno{\text{(B.3)}}
\end{equation*}
where $L_n^{|b|}$ are associated Laguerre polynomials.

In the case $|b|=1/2$, the solutions of Eq.~(B.1) are different. The spectrum is
\begin{equation*}
m_n^2=2n+1+c, \qquad n=0,1,2,\dots,
\eqno{\text{(B.4)}}
\end{equation*}
and the corresponding normalized eigenfunctions are given by
\begin{equation*}
\psi_n=\sqrt{\frac{1}{2^nn!}}\pi^{-1/4}e^{-z^2/2}H_n(z),
\eqno{\text{(B.5)}}
\end{equation*}
where $H_n$ are Hermite polynomials.

In the case $|b|<1/2$, there is no finite discrete spectrum.

\end{document}